\documentclass{amsart}
\usepackage{amsmath}
\usepackage{amssymb}
\usepackage{latexsym}

\newcommand{\ZZ}{{\mathbb Z}}
\newcommand{\RR}{{\mathbb R}}
\newcommand{\CC}{{\mathbb C}}

\newcommand{\QQ}{{\mathbb Q}}

\newtheorem{theorem}{Theorem}   
\newtheorem{lemma}{Lemma}[section]   
\newtheorem{prop}[lemma]{Proposition}

\newcommand{\Bxi}{{\boldsymbol{\xi}}}
\newcommand{\Beta}{{\boldsymbol{\eta}}}
   
\begin{document}
\title{Reflection symmetries of almost periodic functions}
\author[D.~Damanik, R.~Killip]{David Damanik$\,^{1,2}$ and Rowan Killip$\,^{1}$}
\maketitle
\vspace{0.3cm}    
\noindent    
$^1$ Department of Mathematics 253--37, California Institute of Technology,    
Pasadena, CA 91125, U.S.A.\\[0.2cm]    
$^2$ Fachbereich Mathematik, Johann Wolfgang Goethe-Universit\"at,    
60054 Frankfurt, Germany\\[0.3cm]    
\begin{abstract}
We study global reflection symmetries of almost periodic functions. In the non-limit periodic case, we establish an upper bound on the Haar measure of the set of those elements in the hull which are almost symmetric about the origin. As an application of this result we prove that in the non-limit periodic case, the criterion of Jitomirskaya and Simon ensuring absence of eigenvalues for almost periodic Schr\"odinger operators is only applicable on a set of zero Haar measure. We complement this by giving examples of limit periodic functions where the Jitomirskaya-Simon criterion can be applied to every element of the hull.
\end{abstract}
\section{Introduction}
In this article we are interested in global reflection symmetries in almost periodic functions.
Given a bounded function $f\!:\!\ZZ\rightarrow\RR$ we denote the set of translates of $f$ by $U_0$. The function $f$ is said
to be almost periodic if $U_0$ is precompact in $\ell^\infty(\ZZ)$.  The closure of this set is called the hull of $f$ and
will be denoted by $U$.

Let us suppose that $f$ is almost periodic. As we shall see in Section 2, $U$ admits an abelian group structure. We use $\mu$ to denote the Haar measure on this compact abelian group and normalize it by $\mu(U)=1$.

We are be interested in the set of those elements in $U$ which are almost symmetric about the origin.  To state this
more formally, we introduce the reflection map $R\!:\!\ell^\infty(\ZZ)\to\ell^\infty(\ZZ)$, $[Rg](n)=g(-n)$
and the standard metric on $\ell^\infty(\ZZ)$, $d(g,g') = \|g-g'\|_{\ell^\infty(\ZZ)}$. Then we define
$$
U_{{\rm r}}(\varepsilon) = \{ g \in U : d(Rg,g) \le \varepsilon \}
$$
for each $\varepsilon \geq 0$.

There are two further notions which we wish to introduce: The translation operator maps $U$ onto itself by
$[Tg](n) = g(n+1)$. A function is said to be limit periodic if it is the
uniform limit of a sequence of periodic functions.  With this definition we may state

\begin{theorem}\label{refl}
Suppose $f$ is almost periodic but not limit periodic. Then there exists a constant $C$ such that for all $\varepsilon > 0$, $\mu(U_{{\rm r}}(\varepsilon)) \le C \varepsilon$.
\end{theorem}

Our motivation for Theorem \ref{refl} is to understand the applicability of a criterion due to Jitomirskaya and Simon \cite{js}.  This is a criterion for certain Schr\"odinger operators to have no eigenvalues.

To each $g\in U$ one may associate a discrete Schr\"odinger operator
$$
[H_g u](n) = u(n+1) + u(n-1) + g(n) u(n) \quad \text{acting in $\ell^2(\ZZ)$.}
$$
 It is well known that the spectral properties of these operators are $\mu$-almost surely independent of $g$ (in fact, the spectrum and the absolutely continuous spectrum are completely independent of $g$, see Last-Simon \cite{ls}). The most prominent example is the almost Mathieu operator family given by
$$
f(n) = \lambda \cos (2\pi \alpha n),
$$
where $\lambda > 0$ and $\alpha \in (0,1)$ is irrational (see \cite{j1,l} for reviews). Note that $f$ is even and hence $U_{{\rm r}}(0)$ is non-empty. It was conjectured by Aubry and Andre \cite{aa} that for $\lambda > 2$, the operators should exhibit pure point spectrum. However, based on Gordon \cite{g}, Avron and Simon \cite{as2} proved that for Liouville $\alpha$,
every $H_g$ has empty point spectrum.
The conjecture was then modified to assert pure point spectrum for Diophantine $\alpha$. However, Jitomirskaya and Simon \cite{js} proved that if $f$ is even, $\sigma_{{\rm pp}}(H_g) = \emptyset$ for a dense $G_\delta$ of elements $g \in U$.
It is the criterion they developed to prove this that we wish to discuss.
Given $f$ they prescribe a number $B>0$ and introduce sets
$$
U^{(n)} = \{ g \in U : d(RT^{2n}g,g) \le \exp (-Bn) \}
$$
and
$$
U_{{\rm rs}} = \limsup_{n \rightarrow \infty} U^{(n)}.
$$
Jitomirskaya and Simon then prove that
\begin{equation}\label{jsincont}
U_{{\rm rs}} \subseteq \{g \in U : \sigma_{{\rm pp}}(h_g) = \emptyset\}
\end{equation}
and, moreover, that if $f$ is even, then $U_{{\rm rs}}$ is a dense $G_\delta$ subset of $U$.
In view of \eqref{jsincont} it is therefore natural to study $\mu(U_{{\rm rs}})$.

Turning back to the almost Mathieu family, the strongest localization result one can hope for is pure point spectrum (in fact with exponentially decaying eigenfunctions) for every $\lambda > 2$, Lebesgue-almost every $\alpha$ and $\mu$-almost every $g \in U$. Indeed, this modified form of the original Aubry-Andre conjecture has been recently established by Jitomirskaya \cite{j2} (we remark that using duality, one can infer consequences for the case $\lambda < 2$ \cite{gjls,j2}). By \eqref{jsincont} this in turn implies $\mu (U_{{\rm rs}}) = 0$ in the almost Mathieu case, for $\lambda > 2$ and almost every $\alpha$. Theorem \ref{zeromujs} below, which follows quickly from Theorem \ref{refl}, establishes this directly for all non-limit periodic hulls.

\begin{theorem}\label{zeromujs}
Suppose $f$ is almost periodic but not limit periodic. Then, $\mu (U_{{\rm rs}}) = 0$.
\end{theorem}

This proves that, although the Jitomirskaya-Simon criterion may be used to prove absence of eigenvalues for generic $H_g$,
the set of applicability will always have zero Haar measure.
To demonstrate that in general one cannot expect this result for limit periodic $f$, we consider the example 
\begin{equation}\label{counterex}
f(n) = \sum_{k=1}^\infty a_k \cos\! \left(\tfrac{2\pi}{k} n \right)
\end{equation}
with rapidly decaying $a_k$ and show that $U_{{\rm rs}}$ may consist of the entire hull.

\begin{theorem}\label{jseqall}
Let $f$ be given by \eqref{counterex} and suppose that
\begin{equation}\label{decay}
\liminf_{m \rightarrow \infty} \,\, 4 \exp(m!B)\!\!\! \sum_{k=m+1}^\infty |a_k| < 1.
\end{equation}
Then we have $U_{{\rm rs}} = U$.
\end{theorem}

The organization is as follows. In Section 2 we recall some basic notations and observations for almost periodic hulls $U$. Section 3 provides the proofs of Theorems \ref{refl} and \ref{zeromujs}. Finally, Section 4 contains a proof of Theorem \ref{jseqall}.

\section{Preliminaries}
Much of the material presented in this section is an elementary consequence of the theories of almost periodic functions
and of compact abelian groups---both core topics in harmonic analysis.
The appendix of \cite{as} provides an extremely concise development of the relation of these two topics in the case of almost periodic functions on $\RR$. Background material is available in Katznelson \cite{k} and in a more abstract setting in Hewitt-Ross \cite{hr}.

Given a bounded function $f : \ZZ \rightarrow \RR$, we denote the translates of $f$ by $f_m(n) = T^m f (n) = f(n+m)$ and let $U_0 = \{f_m : m \in \ZZ\}$. Recall from the introduction that the function $f$ is called almost periodic if $U_0$ is precompact in $\ell^\infty (\ZZ)$ and that we denote the closure of $U_0$ by $U$.

The set $U_0$ admits a group structure, $[f_m \ast f_{m'}](n) = f_{m+m'}(n)$, which can then be extended to $U$. In particular, $[f_m \ast g](n) = g(n+m)$ for all $g \in U$. Henceforth we shall assume that $f$ is almost periodic so that
$U$ is a compact abelian group and take $\mu$ to be the normalized Haar measure on $U$; $\mu(U) = 1$.

Let $\widehat{U}$ denote the dual group, that is, the group of continuous homomorphisms $\Bxi:U \rightarrow S^1 = \{z \in \CC : |z| = 1\}$.
As $U_0$ is a dense subset of $U$, each character $\Bxi \in \widehat{U}$ is uniquely determined by its values on $U_0$, and hence by $\xi = \Bxi(f_1) \in S^1$. Notice that $\Bxi(f_m) = \xi^m$. We shall use freely the identification of $\widehat{U}$ as a subset of $S^1$ in what follows.

The set $\widehat{U}$ forms an orthonormal basis of $L^2(U;d\mu)$ and so one obtains Fourier expansions for functions on $U$. For example, let $\delta_0:U \rightarrow \RR$ be defined by $\delta_0(g) = g(0)$. Then
$$
\delta_0(g) = \sum \widehat{\delta_0}(\Bxi) \Bxi(g),
$$
where the sum should be interpreted in the $L^2(U;d\mu)$ sense and
$$
\widehat{\delta_0} (\Bxi) = \int_U \delta_0(g) \overline{\Bxi(g)} d\mu(g).
$$
It is also possible to reconstruct each element of $U$ from its character values. To be precise,
\begin{equation}\label{char}
g(n) = \delta_0 ( f_n \ast g ) = \sum \widehat{\delta_0} (\Bxi) \xi^n \Bxi(g),
\end{equation}
where the sum may be made to converge uniformly by employing the summability method of \cite[\S5.18]{k}. One particular case of this is $f(n) = \sum \widehat{\delta_0}(\Bxi) \xi^n$ which shows that $\widehat{\delta_0} (\Bxi)$ are the Fourier coefficients of $f$, more commonly written as $\widehat{f}(\{{\Bxi}\})$ (cf.~\cite{k}).

The set of $\Bxi \in \widehat{U}$ with $\widehat{f}(\{\Bxi\}) \not= 0$ is called the spectrum of $f$ and is denoted by $\sigma(f)$. Since \eqref{char} shows that each $g \in U$ is determined by merely $\Bxi(g)$, $\Bxi \in \sigma(f)$, it should come as no surprise that $\widehat{U}$ is generated by $\sigma(f)$; $\widehat{U} = \langle \sigma(f) \rangle$.

Recall that the function $f$ is limit periodic if there exists a sequence of periodic functions $g^{(n)} : \ZZ \rightarrow \RR$ which converge to $f$ in $\ell^\infty (\ZZ)$. The spectral condition is given by the following proposition (cf.~Theorem A.1.2 of \cite{as}).

\begin{prop}\label{limperchar}
$f$ is limit periodic $\Leftrightarrow$ $\sigma(f) \subseteq \exp\{2\pi i \QQ\}$ $\Leftrightarrow$ $\widehat{U} \subseteq \exp\{2\pi i \QQ\}$.
\end{prop}
\noindent{\it Proof.} If $f$ is limit periodic, then for each $\eta \in \exp\{2 \pi i \RR \setminus \QQ\}$, we have
\begin{equation}\label{zeromean}
\lim_{N \rightarrow \infty} \frac{1}{2N+1} \sum_{n=-N}^N \eta^{-n} f(n) = 0.
\end{equation}
This is easily shown for periodic $f$; it then follows for limit periodic $f$ by uniform approximation. It is also easy to show that
$$
\lim_{N \rightarrow \infty} \frac{1}{2N+1} \sum_{n=-N}^N \eta^{-n} \xi^n = \left\{ \begin{array}{ll} 1, & \eta = \xi\\ 0, & \eta \not= \xi \end{array} \right. .
$$
Applying this and \eqref{zeromean} to \eqref{char}, we see that if $f$ is limit periodic and $\eta \in \sigma(f)$, then $\eta \in \exp\{2 \pi i \QQ\}$. Conversely, if $\sigma(f) \subseteq \exp\{2 \pi i \QQ\}$, then applying summability to \eqref{char} shows that $f$ is a uniform limit of periodic functions. The other equivalence is a direct consequence of $\widehat{U} = \langle \sigma(f) \rangle$.\hfill$\Box$

\section{The measure of $U_{{\rm r}}(\varepsilon)$ and the applicability of the Jitomirskaya-Simon criterion}
In this section we prove Theorems \ref{refl} and \ref{zeromujs}.\\[5mm]
\noindent{\it Proof of Theorem \ref{refl}.} Given $g \in U_{{\rm r}}(\varepsilon)$, define $\Delta_g : U \rightarrow \RR$ by 
$$
\Delta_g(h) = \delta_0(g \ast h) - \delta_0(g \ast h^{-1}).
$$
As $g \in U_{{\rm r}}(\varepsilon)$, we have $|\Delta_g(h)| \le \varepsilon$ for every $h \in U_0$ (recall that $[g \ast f_m](n) = g(n+m)$). Since $U_0$ is dense in $U$ and $\Delta_g$ is continuous, this gives $|\Delta_g(h)| \le \varepsilon$ for every $h \in U$ and hence $\| \Delta_g \|^2_{L^2(U;d\mu)} \le \varepsilon^2$. Consider the Fourier expansion of $\Delta_g$. One finds 
$$
\widehat{\Delta_g} (\Bxi) = \widehat{\delta_0} (\Bxi) \Big[\Bxi(g) - \overline{\Bxi(g)}\Big]
$$
and thus by Parseval,
\begin{equation}\label{eifs}
\sum | \widehat{\delta_0} (\Bxi) |^2 |\Bxi(g) - \overline{\Bxi(g)} |^2 = \| \Delta_g\|^2_{L^2(U;d\mu)} \le \varepsilon^2.
\end{equation}
  Each $\Bxi \in \smash{\widehat{U}}$ is a continuous homomorphism from $U$ to $S^1$ and so each $\Bxi (U)$ is a compact subgroup of $S^1$. Of course, the characters $\Bxi$ for which $\Bxi(U)$ is discrete are precisely those with $\xi \in \exp\{2 \pi i \QQ \}$. Since $f$ is assumed not to be limit periodic, by Proposition \ref{limperchar} there exists $\eta \in \sigma(f)$ which is not an element of $\exp\{2 \pi i \QQ\}$. Consequently, $\Beta(U) = S^1$ and
\begin{equation}\label{etameas}
\mu \big( \{ g \in U : |\Beta(g) - \overline{\Beta(g)}| \leq \alpha \} \big)
	= \big|\{ \theta \in [0,1) : 4 \sin^2(2\pi \theta) \leq \alpha^2\}\big| \leq \frac{\alpha}{2},
\end{equation}
where $|\cdot|$ denotes Lebesgue measure. Also, since $\eta \in \sigma(f)$, $\widehat{\delta_0} (\Beta) = \widehat{f}(\{{\Beta}\}) \not= 0$. Thus we may combine \eqref{eifs} with \eqref{etameas} to obtain
$$
\mu(U_{{\rm r}}(\varepsilon)) \le \frac{\varepsilon}{2 \big| \widehat{f}(\{{\Beta}\}) \big|}.
$$
This proves the theorem.\hfill$\Box$

\vspace{\baselineskip}

\noindent
{\it Proof of Theorem \ref{zeromujs}.} It is easy to check that the maps $R,T$ defined in Section 1 obey $RT = T^{-1}R$. This and the fact that $\mu$ is $T$-invariant yield
$$
\mu\big(\{ g \in U : d(RT^{2n} g,g) \le \exp (-Bn) \}\big) = \mu\big(U_{{\rm r}}(e^{-Bn})\big)\leq Ce^{-Bn}.
$$
The assertion now follows from Borel-Cantelli.\hfill$\Box$

\section{A counterexample}

In this section we consider limit periodic hulls $U$ generated by functions of the form \eqref{counterex} and prove that under assumption \eqref{decay}, every element of the hull has reflection symmetries as required in the definition of $U_{{\rm rs}}$.
\\[5mm]
{\it Proof of Theorem \ref{jseqall}.} Let $f$ be given by \eqref{counterex} and define
$f^{(m)} (n) = \sum_{k=1}^m a_k \cos (\frac{2\pi}{k} n)$.
Obviously, $f^{(m)}$ is even and periodic with period bounded by $(m!)$. Thus $f^{(m)}$ has infinitely many centers of global reflection symmetry and two consecutive centers are separated by a distance of at most $(m!)$. Consider now an arbitrary element $g$ in the hull $U$ of $f$. By assumption there exists a sequence $m_j \rightarrow \infty$ such that $4 \exp(m_j!B)\varepsilon_j < 1$ for every $j$, where $\varepsilon_j = \sum_{k=m_j+1}^\infty |a_k|$.
Now choose some translate $T^{l_j}f$ of $f$ which is $\varepsilon_j$-close to $g$. Then, $T^{l_j} f^{(m_j)}$ is $2\varepsilon_j$-close to $g$ and has at least one center of global symmetry in $\{1,\ldots,m_j!\}$. Thus there exists an $r_j \in \{1,\ldots,m_j!\}$ such that $d(RT^{2r_j} g,g) \le \exp (-m_j!B) \le \exp (-r_j!B)$.\hfill$\Box$\\[5mm] 

\noindent{\it Acknowledgments.} We would like to thank T.~Wolff for useful discussions. D.~D.~was supported by the German Academic Exchange Service through Hochschulsonderprogramm III (Postdoktoranden).  R.~K.~was supported by a Sloan Doctoral Dissertation fellowship.

\end{document}